\documentclass{article}
\usepackage{spconf,amsmath,graphicx}
\usepackage{booktabs}
\usepackage{subfigure}
\usepackage{cite}
\usepackage[urlcolor=blue]{hyperref}

\title{Transformer-S2A: Robust and Efficient Speech-to-Animation}

\name{Liyang Chen$^{1,3}$, Zhiyong Wu$^{1*}$, Jun Ling$^{2,3}$, Runnan Li$^{3*}$ \thanks{*Corresponding authors.}, Xu Tan$^3$, Sheng Zhao$^3$}
\address{Author Affiliation(s)}

\address{
    $^1$ Shenzhen International Graduate School, Tsinghua University, Shenzhen, China\\
    $^2$ Shanghai Jiao Tong University, Shanghai, China\\
    $^3$ Microsoft, China\\
    \small{
        cly21$@$mails.tsinghua.edu.cn, 
        zywu$@$se.cuhk.edu.hk,
        lingjun$@$sjtu.edu.cn,
        \{runnan.li, xu.tan, sheng.zhao\}$@$microsoft.com
    }
}

\begin{document}

\maketitle

\begin{abstract}
We propose a novel robust and efficient Speech-to-Animation (S2A) approach for synchronized facial animation generation in human-computer interaction.  Compared with conventional approaches, the proposed approach utilizes phonetic posteriorgrams (PPGs) of spoken phonemes as input to ensure the cross-language and cross-speaker ability, and introduces corresponding prosody features (i.e. pitch and energy) to further enhance the expression of generated animation. Mixture-of-experts (MOE)-based Transformer is employed to better model contextual information while provide significant optimization on computation efficiency.  Experiments demonstrate the effectiveness of the proposed approach on both objective and subjective evaluation with 17$\times$ inference speedup compared with the state-of-the-art approach.


\end{abstract}

\section{Introduction}
\label{sec:intro}

With the development of human-computer interaction, the single modality of interaction can less and less satisfy the various interaction scenarios. To enhance user experience, especially in conversational scenarios, such as education and dialogue, multi-modal interaction is thus becoming more and more important. In these modalities, speech and vision, the two most natural communication channels, have drawn rising attraction. Since latest text-to-speech technology \cite{fastspeech2} can already synthesize human-like speech, how to produce high-quality synchronized facial animation becomes the key to provide high-performance multi-modal service.  Speech-to-animation (S2A) technology is then developed to address this task. The generated animation parameters from S2A can be any predefined visual parameters, such as blendshapes or landmarks, to drive virtual avatars in different communication scenarios. S2A can also help in conventional content generation scenarios, such as movie production process, to minimize the expensive, time-consuming and non-scalable effort in producing synchronized facial animations.

S2A technology is developed to automatically estimate synchronized facial animation parameters from given speech, including lip, jaw and cheek-area movements. A rendering engine like Unreal Engine 4 (UE4) is then employed to generate the final animated avatar with these predicted parameters. One major research interest in S2A development is to find out appropriate speech representation for building a robust S2A system \cite{netease, audioDVP, VOCA, ppgArxiv, NVP}.
Compared with mel-frequency cepstral coefficients (MFCC) used in \cite{netease, audioDVP}, the latest proposed phonetic posteriorgrams (PPGs)-based approaches \cite{VOCA, ppgArxiv, NVP} can leverage the property of pre-trained large-scale automatic speech recognition (ASR) model \cite{deepspeech}, to provide speaker and language independent abilities for S2A systems. However, since PPGs are more about linguistic information, using PPGs only as input always fails in preserving prosody information and results in poor expressiveness in animation generation.

The model construction is another key component in S2A development. Early attempts about the unit selection \cite{unitSel1,unitSel2} were made to map acoustic phonemes to visual visemes. These methods suffer from poor synchronization between speech and animation, since they fail to embody continuous variance of animations in adjacent frames.
With the development of deep learning, data-driven approaches including frame-level \cite{NVP, audioDVP} and sequence-level \cite{ppgArxiv, netease, TMM2021} methods are developed for S2A.  In frame-level methods, $T$ frames of features in a temporal window are stacked as input, and the animations are predicted with the
window sliding along the given speech sequence. These methods can maintain temporal coherency in modeling but ignore the dependencies between suprasegmentals, leading to unexpected lip tremble since the preceding and following phonemes can significantly affect the animation of the current phoneme spoken.  The sequence-level methods \cite{netease, ppgArxiv, TMM2021} directly process the whole speech sequence instead of batching it into multiple frames. 
A typical implementation is to employ bi-directional long short-term memory (BLSTM) to explore time-related features and generate plausible facial animations.
These methods still have limitation in modeling suprasegmental contextual information and the auto-regressive manner could limit the inference efficiency.

In this paper, we propose a robust and efficient S2A system to address aforementioned issues by using additional prosody features and mixture-of-experts (MOE) Transformer. For prosody features, we introduce pitch and energy as inputs to provide more variation information in speech.  We further incorporate MOE-based Transformer to better model the contextual information and speed up the inference speed of sequential model. The contribution can be summarized as:
\begin{itemize}
	\item Propose a robust and efficient S2A system to generate facial animation from speech. The proposed system is speaker and language independent with the use of PPG as input, and the expression of generated animation is better than the state-of-the-art (SOTA) with the use of prosody features as additional inputs.
	\item Propose MOE-based Transformer for S2A model construction. With the MOE layer, the system can better exploit the contextual information from the given input sequence by automatically selecting expert candidates.
\end{itemize}

\section{Methodology}
\label{sec:methodology}
In this section, we firstly describe the optimized input features and 
then introduce MOE-based Transformer for modeling.

\subsection{PPG Augmented with Prosody Features}

Since collecting paired speech-animation data is a time-consuming and expensive work,
it is challenging to develop a speaker-independent and cross-lingual S2A system. Inspired by previous works \cite{VOCA, ppg2016, ppgArxiv},
we utilize PPGs as the basic speech representation to generate animations across speakers 
and languages with comparable synthesized quality.

\begin{figure}[h]
	\centering
	\includegraphics[scale=0.38]{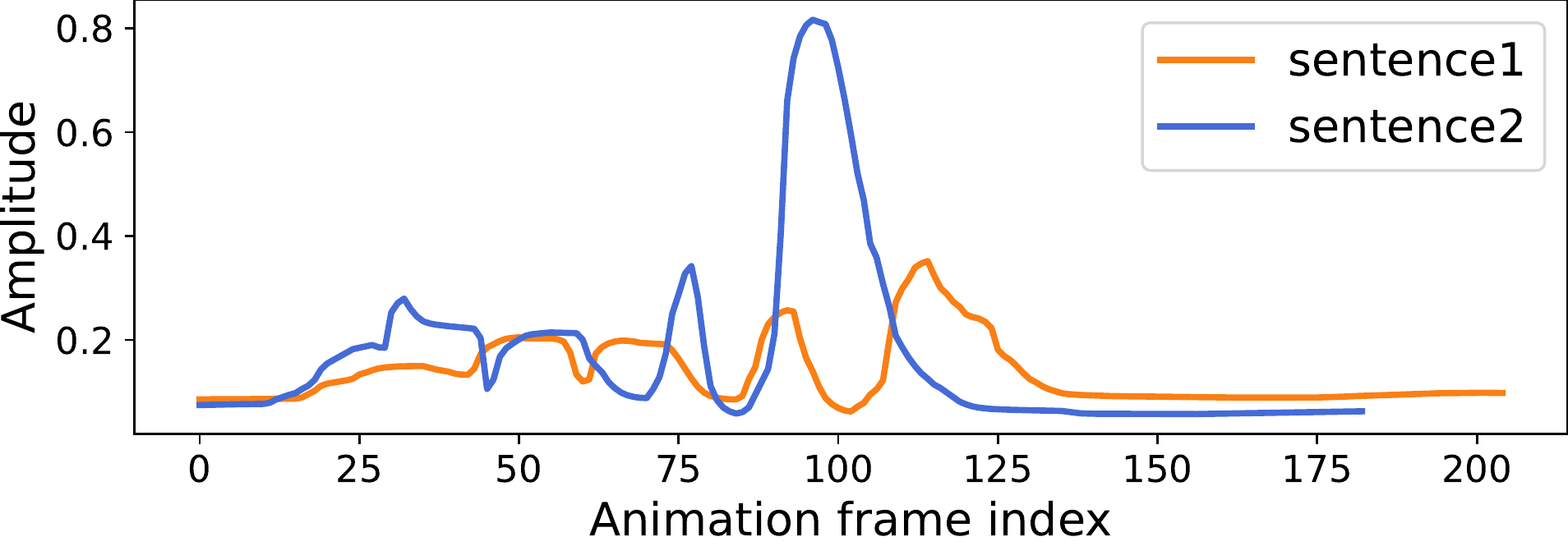}
	\caption{The jawOpen curves of two sentences with the same context ``millions of dollars".}
	\label{fig:animations}
\end{figure}

PPGs conveys more linguistic information but less acoustic information, since prosody related information has been removed in the extraction of PPGs. Considering PPGs from the same speech transcription may be assigned with various animation presentations, the lack of prosody information may result in poor expression and variance in generated animations, which refers to the typical one-to-many mapping issue\cite{fastspeech2, imgTrans}. As shown in Figure \ref{fig:animations}, the curves of jawOpen, one of the crucial parameters in animations, differ a lot from each other but share the same speech content.  To prevent the S2A model failing in capturing animation variations, additional prosody features, pitch and energy, are introduced to provide clear guidance for the model. It is consistent with the common sense that the louder and higher-pitched voice always makes the wider mouth open.

Spectral features \cite{lipSync3D,netease} are not directly used in this work for two reasons.
1) The spectral features are speaker-dependent. 
2) Empirically, redundant information (e.g. background noise) in spectral features that has no relationship with pronunciation disturbs the training and causes instability in generation. By means of PPGs augmented with the prosody features, the proposed approach obtains noise, speaker and language auto-adapted ability in a more data-efficient way.

\subsection{MOE-based Transformer}
\begin{figure}
	\centering
	\includegraphics[scale=0.7]{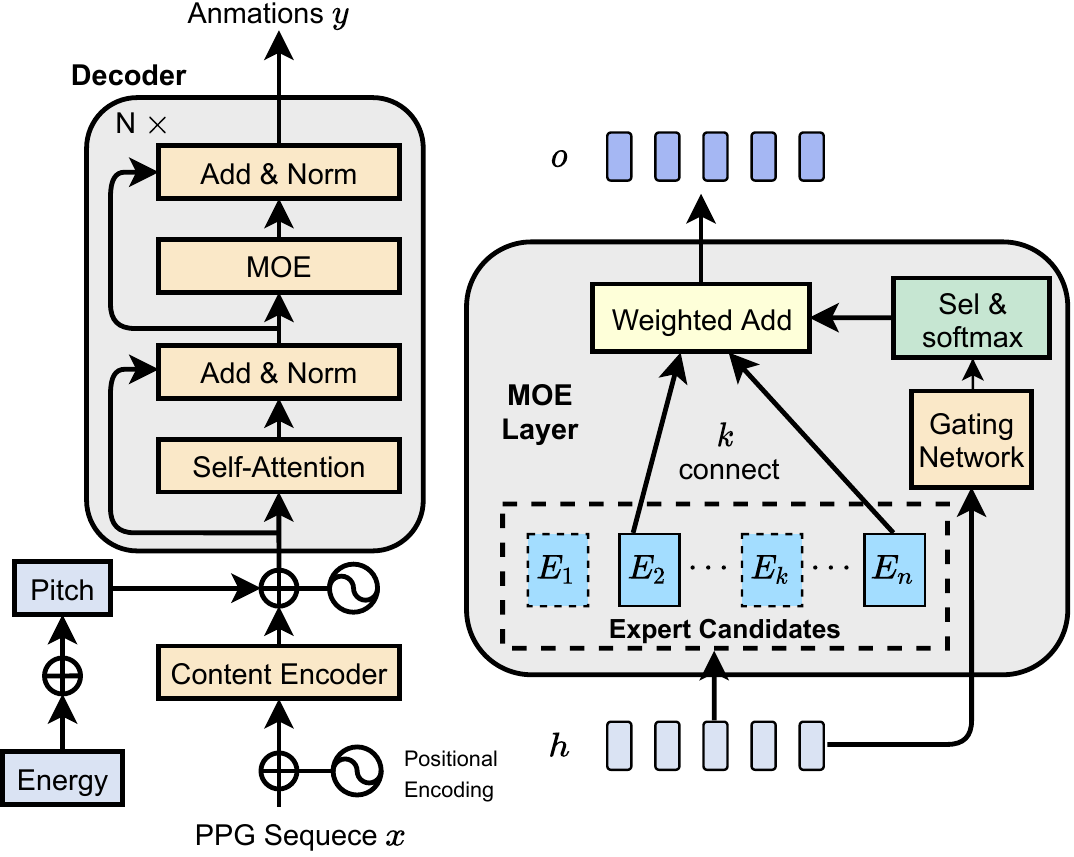}
	\caption{Architecture of Transformer-S2A. Prosody features are concatenated while positional encoding is added to PPGs.}
	\label{fig:MOE}
\end{figure}

Transformer \cite{Transformer}, a typical sequence-to-sequence network, has achieved great success in machine translation \cite{bert} and speech processing \cite{fastspeech2}. It mainly consists of two components: the self-attention mechanism and the feed-forward network. Compared with RNN-based models, Transformer generates outputs without conditioning current input on the previous output, and thus enables parallel training. In the meantime, the self-attention captures long-range and multi-level dependencies across the whole sequence by presenting the global context to each input frame. Nevertheless, the many-to-many mapping problem \cite{unitSel1} as well as the cross-modal nature of S2A demands for specific architectures with more powerful context extraction ability, different from those employed for language understanding with single-modal.

The architecture of the proposed model is shown in Figure \ref{fig:MOE}, comprising a content encoder and an expert decoder.
Let $ x=(x_1,\cdots	, x_{t1})$ be the sequence of input PPGs, and $y=(y_1,\cdots, y_{t2})$ be the sequence of target
facial animations. Due to the different sampling rate of PPGs and animations, we first interpolate $x$ to become the 
same length $t2$ with y \cite{VOCA}. The content encoder that is stacked with the blocks in the Transformer encoder \cite{Transformer}, takes PPGs
as inputs and produces the hidden representation. The goal of the encoder is to extract robust text-related information
and model the co-articulation effects from PPGs.

The expert decoder is constructed with several identical Transformer blocks that based on the MOE 
layer \cite{MOE, fastMOE}. As illustrated in Figure \ref{fig:MOE}, we concatenate the hidden representation
and prosody features pitch and energy to form the input sequence to the decoder.
After being processed by the self-attention layer, the sequence $h$ is further passed into the MOE layer, which consists
of multiple experts $E_1,\cdots,E_n$ and a gating network $G$. The experts are of the same architecture but separated parameters, 
each comprising a convolutional layer with stride 1 and a fully connected layer. The gating network gives scores on
how much effort each expert may contribute over the input. Only the experts with the top-k values are kept, and the
$softmax$ function is then applied over the experts. 
\begin{align}
Prob(h) = Softmax(Score(h))
\end{align}
\begin{align}
	Score(h)_i = \left\{
	\begin{array}{ll} 
	G_i(h) ,  &\text{for top $k$ elements in $G(h)$} \\
	-\infty,  &\text{otherwise}.
	\end{array}	\right.
\end{align}
The output of MOE layer can be formulated as 
\begin{align}
	o = \sum_{i=0}^{n}E_i(h)Prob_i(h)
\end{align}

The number of expert candidates is set as 48 which roughly equals to the number of distinct  phonemes in Mandarin or English, and $k$ is set as 16 according to the  defined basic visemes in \cite{wu2006}. Although MOE is originally designed for enlarging the model capacity without proportionally increasing computation \cite{MOE}, we find it works for S2A task under such setting. This can be attributed to the inductive bias for phoneme-to-viseme introduced by the MOE structure, where all the experts collaborate to search the specific mapping from the hypothesis space in a more efficient way.

\section{Experiments}
\label{sec:pagestyle}
\subsection{Experimental Setup}
\textbf{Rendering engine.} In this work, the digital avatar character is developed based on UE4, that is, 
a game creation engine commonly used in the gaming industry. Benefiting from the UE creation nature that the animation is disentangled
with the character, the animation predicted by the pre-trained S2A model is able to drive any digital character that is designed following
the developing rule of UE4. Hence, these features contribute to a reliable character-independent pipeline while almost no prior knowledge
about game creation is required. This creation tool also can be replaced with other 3D computer graphics application, e.g. Unity and Maya. Synthesized demo videos are available online\footnote{https://thuhcsi.github.io/icassp2022-Transformer-S2A}. \\
\textbf{Data collection.}
We utilize the mobile software Live Link Face\footnote{https://apps.apple.com/us/app/live-link-face/id1495370836} and 
the 3D character kite-boy\footnote{https://www.unrealengine.com/marketplace/en-US/product/kite-demo}
that is freely available for non-commercial use to collect training data. The blendshape coefficients recorded by the software contain timestamps of dictionary that describe the facial expression of the detected face, and we only select 32 dimensional coefficients that related to pronunciation as the target animation for training.
Compared with other works \cite{netease, DurIAN} that involve commercial system or extra depth estimation algorithm, the only thing we need is an iPhone that supports face recognition, which greatly reduces the complexity and cost in high-quality data collection. Only \textbf{one-hour} paired speech and animation data is collected from a single person for model training. Ten-minute outside data is reserved as the test set.\\
\textbf{PPG extraction.}
we pre-train an automatic speech recognition (ASR) model with easily-available speech corpora of thousands of Chinese and English speakers, 
and extract the bottleneck of the model with dimension 64 as PPG. Speech data is sampled at 16 kHz and 80-dimensional filter banks computed with a sliding hamming window of 40ms width and 20ms overlap are used as the inputs to the ASR model. \\
\textbf{Data pre-process for S2A.}
The prosody features pitch and energy are extracted with the same setting with filter banks in the PPG model. Voice activity detection is performed to remove silence frames at the start and end portion of the utterance. In particular, data normalization on the facial animation is vital to synthesize smoother lip movements.

\subsection{Evaluations}
\subsubsection{Objective Evaluation}
To evaluate the performance of the proposed approach, we conduct quantitative evaluation by calculating root mean square error (RMSE) between the predicted animations and ground-truth. The RMSEs on the entire animations and two crucial animations (jawOpen and mouthClose) are separately described.
We compare our approach with 1) NVP \cite{NVP}, the state-of-the-art (SOTA) code-available approach in photo-realistic talking face
generation task (using the implementation of Audio2ExpressionNet\footnote{https://github.com/keetsky/NeuralVoicePuppetry}, which
predicts facial expression in frame level)
2) BLSTM, which has been adopted by several S2A systems \cite{netease, ppgArxiv, TMM2021}. 

As described in Table \ref{tab:RMSE}, the RMSE of the entire animation decreases with the sliding window widening, which indicates that more contextual information does lead to more accurate facial movements.
Though BLSTM-based model can capture long-term correlation in sequence level, the synthesis of each animation frame fails to attend the relevant speech frames while NVP imposes manual constraint about attentive region with the sliding window. The proposed model obtains lowest RMSE on both entire and crucial animations.
\begin{table}[t]
	\centering
	\caption{RMSE of different methods. The inputs are the same for all methods.
		The values are divided by 100 for simplicity. P: pitch. E: energy. winLen: the length of the window.}
	\begin{tabular}{lcc}
		\toprule
		Method & Entire & Crucial \\
		\midrule
		NVP winLen=3 & 4.048 $\pm$ 0.640 & 7.951 $\pm$ 1.242 \\
		NVP winLen=9 & 3.899 $\pm$ 0.397 & 8.121 $\pm$ 1.445 \\
		BLSTM & 3.981 $\pm$ 0.510 & 8.185 $\pm$ 1.257 \\
		Proposed & \textbf{3.886 $\pm$ 0.458} & \textbf{7.378 $\pm$ 1.231} \\
		Proposed w/o MOE & 3.930 $\pm$ 0.387 & 7.709 $\pm$ 1.258 \\
		Proposed MFCC & 4.240 $\pm$ 0.712 & 8.013 $\pm$ 1.565 \\
		Proposed w/o P\&E & 3.891 $\pm$ 0.424 & 7.490 $\pm$ 1.230 \\
		\bottomrule
	\end{tabular}%
	\label{tab:RMSE}
\end{table}%

The ablation study is conducted to verify the effectiveness of the modules.
Without MOE layer, RMSE on the crucial animations rises due to the poor model generalization. When using MFCC instead of PPGs as input, significant performance drop can be observed. Removing pitch and energy would also results in performance drop, which demonstrates the effectiveness of prosody features. This phenomenon tends to be more evident if the S2A model is trained on a more expressive dataset.

\begin{figure}
	\centering
	\includegraphics[scale=0.36]{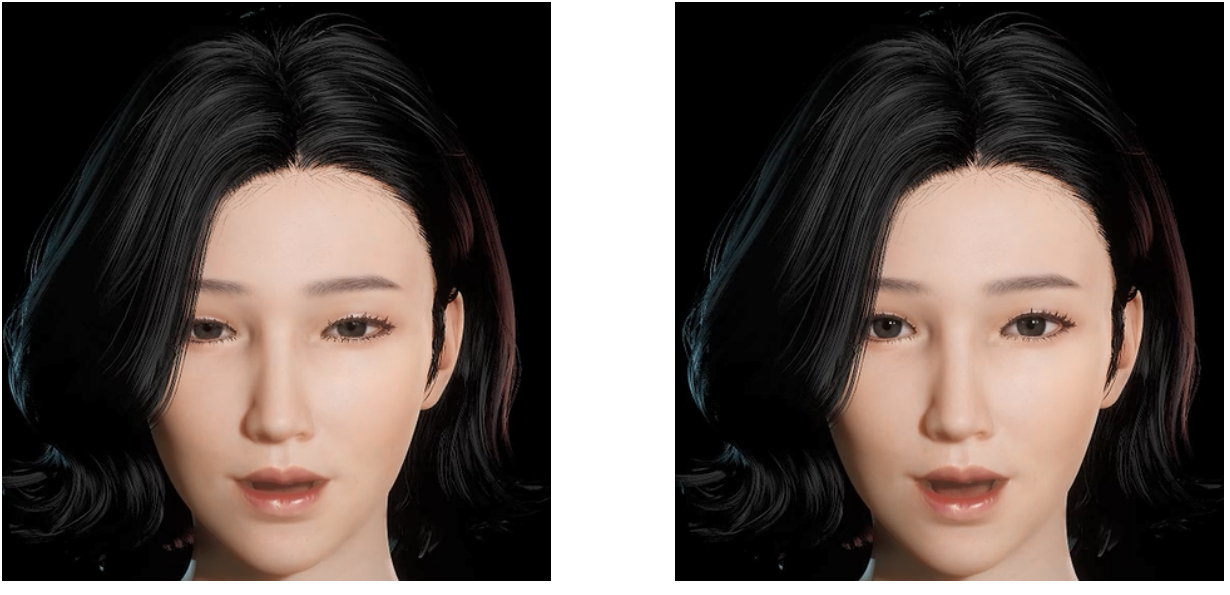}
	\caption{Rendered images with animations predicted from NVP (left) and the proposed (right) after phoneme /i/.}
	\label{fig:render}
\end{figure}

A detailed comparison is shown in Figure \ref{fig:render}. For rendered image frames from NVP (winLen=9) and the proposed approach after speaking Mandarin phoneme /i/, the upper lip in NVP trembles towards the middle position, while the mouth transits to the next phoneme with more expressive and smooth movements in the proposed approach. Similar patterns can be widely observed in the evaluation.

\begin{table}[t]
	\centering
	\caption{The comparison of inference RTF. The evaluation is conducted on a single NVIDIA P100 GPU. RTF denotes the average time to generate one-second animations.}
	\begin{tabular}{lcc}
		\toprule
		Method & RTF & \multicolumn{1}{l}{Speedup} \\
		\midrule
		BLSTM & 0.09486 $\pm$ 0.02077s & 1.00 $\times$ \\
		NVP winLen=9 & \textbf{0.00049 $\pm$ 0.00031s} & \textbf{193.59 $\times$} \\
		Proposed & 0.00534 $\pm$ 0.00219s & 17.76 $\times$ \\
		\bottomrule
	\end{tabular}%
	\label{tab:eff}%
\end{table}%

\subsubsection{Inference Efficiency}
As shown in Table \ref{tab:eff}, we compare the average inference real time factor (RTF) of the proposed model with that of NVP \cite{NVP} and BLSTM \cite{netease, ppgArxiv,TMM2021}. Each utterance for testing is around 12 seconds. The proposed model speeds up the animation synthesis by 17$\times$, compared with BLSTM model. In regard to the frame-level model, our model achieves acceptable speed decline but obtains animations of higher quality.

\subsubsection{Subjective Evaluation}
To further evaluate the perceptual quality of the proposed and the SOTA method when transferring across speakers and languages, ABX test is carried.  Thirty-five judges with proficient English ability are invited to choose the better video in terms of lip-speech synchronization and naturalness. Judges are informed to focus on speech-to-animation performance, and to ignore the difference between characters.  LipSync3D \cite{lipSync3D} (not open-source) is selected as the SOTA method to be compared, and its official demo video \footnote{https://www.youtube.com/watch?v=L1StbX9OznY} is reused in the ABX test directly. Audio from the demo video is extracted as input to the proposed approach for driving a 3D avatar.

\begin{figure}[t]
	\centering
	\includegraphics[scale=0.24]{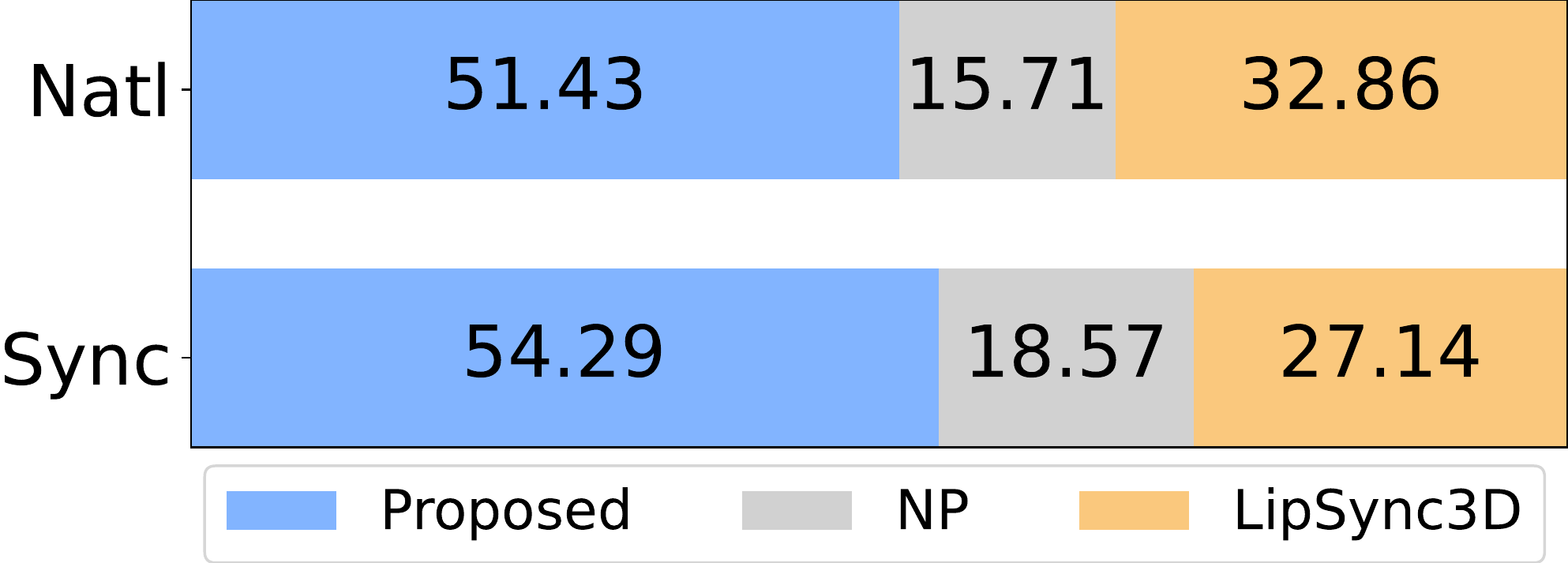}
	\caption{ABX Results. NP: no preference. Sync: lip-speech synchronization. Natl: naturalness.}
	\label{fig:ABX}
\end{figure}

Evaluation result of ABX is presented in Figure \ref{fig:ABX}. The proposed system has outperformed the baseline in both lip-speech synchronization and naturalness. Specially, the proposed S2A model is trained on the Mandarin dataset only and none English data is used, while LipSync3D is directly trained on English datasets.

\section{Conclusion}
\label{sec:typestyle}

In this paper, we propose a robust and efficient S2A system to generate high-quality facial animation from given speech. PPG and prosody features are introduced as inputs to ensure the cross-speaker and cross-language ability while maintain variants in speech. MOE-based Transformer is introduced to better model contextual information while provide significant optimization on computation efficiency.  Experiments demonstrate the effectiveness of the proposed in both quality and efficiency evaluations.

\textbf{Acknowledgement}: This work is supported by National Natural Science Foundation of China (NSFC) (62076144). We would like to thank Digital Domain for their great support in creating digital avatar and providing rendering demos.

\vfill\pagebreak

\bibliographystyle{IEEEbib}
\bibliography{refs}

\end{document}